\documentclass{article}
\usepackage{graphicx}
\usepackage{amsmath}

\input{tcilatex}

\begin{document}

\begin{center}
%\title{Effective spring constants for the elasticaly coupled insertions in
%membranes. Influences of the boundary conditions}
%\maketitle
{\LARGE \textbf{Effective spring constants for the elastically coupled
insertions in membranes.}} \vspace{0.2in}

Yury S. Neustadt$^{1}$ and Michael B. Partenskii$^{2}$

$^{1}$Samara State Academy of Architecture and Civil Engineering, Samara,
443001, Russia\\[0pt]
$^{2}$Department of Chemistry, Brandeis University, Waltham, MA, 02454, USA
\end{center}

\section{Introduction}

Membrane-spanning peptides such as the ion channel gramicidin dimer, cause
membrane deformation which contribute significantly to both the energetics
of the insertion and the membrane-mediated interaction between the channels.
Recently, interest in this field shifted from study of individual channels
to study of collective effects in channel kinetics. Here we discuss an
efficient way to describe the membrane-mediated interaction between the
insertions in terms of coupled harmonic oscillators and corresponding
"spring constants".

\section{Variational principle, boundary conditions and Euler-Lagrange
equation.}

We consider an elastic system extended in $x,y-$plane, where the
deformation\ can be described by a $2-$dimensional field of
\textquotedblright vertical\textquotedblright\ displacement $u(\mathbf{r})$
\ ($\mathbf{r=(}x,y\mathbf{)}$ is the radius vector in the mid-plane of the
system). The examples include smectic and similar models for lipid bilayers
, and \textquotedblright floating plate\textquotedblright\ model of
classical elastic theory (see~\cite{JorMilPar2002} for review).

The elastic boundary problem can be formulated as a variational (minimum)
principle for the energy functional

\begin{equation}
F^{(2)}[u]=\int g^{(2)}(u,\nabla u,\Delta u,...)df  \label{energy}
\end{equation}

where $g^{(2)}$, the surface density of the elastic energy, is a quadratic
function of the surface displacement $u$ and its derivatives. We will
consider a membrane with $N$ cylindrical insertions, assuming that on the
contour $\gamma _{i}$ of $i-th$ insertion both $u(\mathbf{r})$ \ and $\nabla
u(\mathbf{r})$ are fixed functions\ \ of $\mathbf{r}_{\gamma _{i}}$\textbf{, 
}the position vector for the points belonging to $\gamma _{i}$. It leads to
the boundary conditions

\begin{equation}
u(\mathbf{r})|_{\mathbf{r}_{\gamma _{i}}}=u_{i}(\mathbf{r}_{\gamma _{i}})
\label{gu_u_ins}
\end{equation}%
\begin{equation}
\mathbf{\nabla }u(\mathbf{r})|_{\mathbf{r}_{\gamma _{i}}}^{n}=s_{i}(\mathbf{r%
}_{\gamma _{i}})  \label{gu_s_ins}
\end{equation}

\textquotedblright $n$\textquotedblright\ designates the direction normal to 
$\gamma _{i}\,$at the point $\mathbf{r}_{\gamma _{i}}.$Note that the
vertical displacement $u_{i}$ in the immediate contact of a membrane with an
inserted peptide is typically described by the "hydrophobic matching
condition" \cite{Hua86,NieGouAnd98,JorMilPar2002} leading to a particular
case of Eq. \ref{gu_u_ins} with $u_{i}(\mathbf{r}_{\gamma _{i}})=u_{0}=const$%
.

Additional conditions on the external membrane boundary (designated as $%
\gamma _{\infty }$) are:

\begin{equation}
u(\mathbf{r})|_{\gamma _{\infty }}=0  \label{gu_u_inf}
\end{equation}%
\begin{equation}
\mathbf{\nabla }u(\mathbf{r})|_{\gamma _{\infty }}=0  \label{gu_s_inf}
\end{equation}

In many cases, including the biharmonic problem with smooth and continuous
boundaries that we consider (\cite{Cia79}, Ch. 6), the variational principle 
$\delta F^{(2)}=0$ (the minimum condition for the energy functional) leads
to the Euler-Lagrange equation which we present for now as 
\begin{equation}
\emph{L}(u)=0  \label{EL_L}
\end{equation}

\bigskip where $\emph{L}$ is a $\emph{linear}$ differential operator 
\footnote{%
Which means that $\emph{L}\cdot (c_{1}\,u_{1}+c_{2}\,u_{2})=c_{1}\,\emph{L}%
\,\cdot u_{1}+c_{2}\,\emph{L}\cdot u_{2}\;$%
\par
($c_{1}$ and $\ c_{2}$ are the arbitrary constants). The linearity does not
impose, however, any restrictions on the order of the differential equation
(which for most of the applications considered is biquadratic).}. The
elastic energy $E$ $=\min F^{(2)}[u]$ is the value of $F^{(2}[u]$ calculated
with the solutions of Eqs. \ref{EL_L}-\ref{gu_s_inf} in place of $u$. We
will show now that in some important cases $E\ $can be explicitly presented
as a quadratic form of the boundary parameters, such as $u_{i}$ and $\mathbf{%
s}_{i}$.

\section{\protect\bigskip Effective spring constants}

\subsection{\protect\bigskip\ Boundary displacements and contact slopes
fixed to constants}

Suppose now that the boundary displacements and the contact slopes are fixed
at the $i-th$ insertion to the constants

$u_{i}$ and $\ s_{i}.$ Some preliminary results for this case were reported
in \cite{ParJor2002}. Eqs. \ref{gu_u_ins}, \ref{gu_s_ins} can be written as 
\begin{equation}
u(\mathbf{r})|_{\gamma _{i}}=u_{i}  \label{gu_u_const}
\end{equation}%
\begin{equation}
\mathbf{\nabla }u(\mathbf{r})|_{\gamma _{i}}^{n}=s_{i}  \label{gu_s_const}
\end{equation}

We now introduce the \textquotedblright superfinite\textquotedblright\
elements, $\phi _{i}^{u}(\mathbf{r})$ and $\phi _{i}^{s}(\mathbf{r),\,}$%
solutions of Eq. \ref{EL_L} satisfying boundary conditions Eq. \ref{gu_u_inf}%
-\ref{gu_s_inf} and following conditions at the internal boundaries:

\begin{equation}
\phi _{i}^{u}(\mathbf{r})|_{\gamma _{k}}=\delta _{ik},\;\nabla \phi _{i}^{u}(%
\mathbf{r})|_{\gamma _{k}}^{n}=0\;  \label{sfe_GU1}
\end{equation}

\begin{equation}
\phi _{i}^{s}(\mathbf{r})|_{\gamma _{k}}=0,\;\;\nabla \phi _{i}^{s}(\mathbf{r%
})|_{\gamma _{k}}^{n}=\delta _{ik}  \label{sfe_GU2}
\end{equation}

where $\delta _{ik}$ is the Kronecker symbol. The solution of Eqs. \ref{EL_L}
-\ref{gu_s_inf}, \ref{gu_u_const}, \ref{gu_s_const} can be written as a
linear combination of the superfinite elements: 
\begin{equation}
u(\mathbf{r)=}\sum_{i=1}^{N}(u_{i}\phi _{i}^{u}(\mathbf{r)+}s_{i}\phi
_{i}^{s}(\mathbf{r))}  \label{u_2par}
\end{equation}

Substituting this result into Eq. \ref{energy} allows presenting $E$ as a
quadratic form of the boundary parameters:

\begin{equation}
E=\sum_{i=1}^{N}\sum_{j=i}^{N}c_{ij}^{\alpha \beta }\alpha _{i}\beta _{j}
\label{En_quadra}
\end{equation}

where $c_{ij}^{\alpha \beta }$ are the effective spring constants describing
interaction between the insertions $i$ and $j$ ($c_{ii}^{\alpha \beta }$
corresponds to the "self energy"of the $i-$th insertions). Such a "linear
spring model" was first introduced for a single insertion in \cite%
{NieGouAnd98} and later generalized in \cite{ParJor2002}. Eq. \ref{En_quadra}
implies that the additional summation is performed over the repeated
indexes, $\alpha ,\beta \ (=u,s).$

\bigskip The effective spring constants satisfy the symmetry relation 
\begin{equation}
c_{ij}^{\alpha \beta }=c_{ji}^{\beta \alpha }\text{and, consequently}%
,c_{ij}^{\alpha \alpha }=c_{ji}^{\alpha \alpha }.  \label{symmetry}
\end{equation}

As an illustration, consider the expression for $g^{(2)}\;$which is typical
in study of membranes:

\begin{equation}
g^{(2)}=B\;(\Delta u)^{2}+A\;u^{2}  \label{g_2}
\end{equation}

\bigskip where $A\;$and $B$ are proportional respectively to the membrane
stretching and bending elastic constants and can be dependent on $\mathbf{r}$%
, but not on $u;$ $\Delta =\partial ^{2}/\partial x^{2}+\partial
^{2}/\partial y^{2}$ is the Laplace operator.

Combining Eqs. \ref{u_2par}, \ref{g_2} and \ref{En_quadra}, we find the
spring constants: 
\begin{equation}
c_{ij}^{\alpha \beta }=(2-\delta _{ij})\int (B\Delta \phi _{i}^{\alpha
}\Delta \phi _{j}^{\beta }+A\phi _{i}^{\alpha }\phi _{j}^{\beta })df
\label{Spring_c}
\end{equation}

\subsection{Azimuthal variation of the contact slope}

\bigskip

\subsubsection{General formulas\newline
}

A possibility that the contact slope can become anisotropic at the contours
of two interacting insertions was studied in \cite{ParMilJor2002}. The slope
was presented as \ 
\begin{equation}
s_{i}(\mathbf{r}_{\gamma _{i}})=a_{i}+b_{i}f_{i}(\mathbf{r}_{\gamma _{i}})
\label{slope_anis}
\end{equation}

where $\ a_{i}$ and $b_{i}$ do not depend on $\mathbf{r,}$ and $\;f_{i}(%
\mathbf{r})$ are the fixed functions. In practice, functions $f_{i}(\mathbf{r%
})$ approximate the azimuthal variation of the contact slope. If energy is
minimized over the slope parameters $a_{i}$ and $b_{i}$, then different
choices of $\ f_{i}(\mathbf{r})$ result in different families of trial
functions $s_{i}(\mathbf{r}_{\gamma _{i}})\;\cite{ParMilJor2002}$.

The surface displacement $u(\mathbf{r})$ can be expressed now through the
boundary parameters $a_{i},\,b_{i}\,$and $u_{i}.\;$

\begin{equation}
u(\mathbf{r)=}\sum_{i=1}^{N}[u_{i}\phi _{i}^{u}(\mathbf{r)+}a_{i}\phi
_{i}^{a}(\mathbf{r)+}b_{i}\phi _{i}^{b}(\mathbf{r)]}  \label{u_3par}
\end{equation}

where $\phi _{i}^{u}(\mathbf{r)\;}$and\textbf{\ }$\phi _{i}^{a}(\mathbf{r)\;}
$satisfy respectively the boundary conditions \ref{sfe_GU1} and \ref{sfe_GU2}%
, while $\phi _{i}^{b}(\mathbf{r)}$ satisfies the following conditions:

\begin{equation}
\phi _{i}^{b}(\mathbf{r)}|_{\gamma _{k}}=0,\;\;\nabla \phi _{i}^{b}(\mathbf{%
r)}|_{\gamma _{k}}^{n}=\delta _{ik}f_{i}(\mathbf{r}_{\gamma _{i}})
\end{equation}

The energy is still described by Eq. \ref{En_quadra} with the spring
constants defined by Eq. \ref{En_quadra}; \ the additional summation is now
performed over the repeated indexes $\alpha ,\beta =(u,a,b)$ .\newline
\newline

\subsubsection{Applications for two insertions\newline
\newline
}

We consider now two identical insertions. Due to possible fluctuations,
parameters $u_{i},\ a_{i}\;$\ and $b_{i},$ and functions $f_{i}(\mathbf{r}%
)\; $\ for two insertions can still be different. The energy of two
insertions is

\begin{eqnarray}
E
&=&c_{11}^{uu}(u_{1}^{2}+u_{2}^{2})+c_{11}^{aa}(a_{1}^{2}+a_{2}^{2})+c_{11}^{bb}(b_{1}^{2}+b_{2}^{2})\;+2c_{11}^{ua}(u_{1}a_{1}+u_{2}a_{2})+\ \smallskip
\notag \\
&&2c_{11}^{ub}(u_{1}b_{1}+u_{2}b_{2})+\ 2c_{11}^{ab}(a_{1}b_{1}+a_{2}b_{2})+%
\newline
c_{12}^{uu}u_{1}u_{2}+c_{12}^{aa}a_{1}a_{2}+c_{12}^{bb}b_{1}b_{2}+  \notag \\
&&c_{12}^{ua}(u_{1}a_{2}+u_{2}a_{1})+c_{12}^{ub}(u_{1}b_{2}+u_{2}b_{1})+c_{12}^{ab}(a_{1}b_{2}+a_{2}b_{1})
\label{Eq_2}
\end{eqnarray}%
\linebreak where we used the symmetry conditions $c_{11}^{\alpha \beta
}=c_{22}^{\alpha \beta },~\;c_{ij}^{\alpha \beta }=c_{ji}^{\alpha \beta
}=c_{ij}^{\beta \alpha }.$

We now consider equilibrium setting. This case was discussed in \cite%
{ParMilJor2002}. The allowed functions $f_{i}(\mathbf{r})\;$should satisfy
the condition $f_{1}(\mathbf{r})=f_{2}(-\mathbf{r}),$ where $\mathbf{r=}0$
designates the midpoint between the insertions, with boundary parameters for
both channels identical: $u_{1}=u=u,\;a_{1}=a_{2}=a,\;b_{1}=b_{2}=b.~$Then,
the energy can be presented as

\begin{equation}
E=C_{uu}u^{2}\ +C_{aa}a^{2}+C_{bb}b^{2}+C_{ua}ua+C_{ub}ub+C_{ab}ab
\label{Eq_2_same}
\end{equation}

where

\begin{equation}
C_{a\beta }=(2-\delta _{\alpha \beta })(2c_{11}^{\alpha \beta
}+c_{12}^{\alpha \beta })  \label{Spring_same}
\end{equation}

Thus the total number of the effective spring constants is reduced to six.
We intent to study the optimized ("relaxed", "equilibrium") slope, so that%
\begin{equation}
\frac{\partial E}{\partial a}=0,\ \frac{\partial E}{\partial b}=0
\label{Min_cond}
\end{equation}

\bigskip These conditions lead to%
\begin{equation}
a=\frac{2C_{bb}C_{ua}-C_{ab}C_{ub}}{\Delta }u,\;b=\frac{%
2C_{aa}C_{ub}-C_{ab}C_{ua}}{\Delta }u,\;\Delta =C_{ab}^{2}-4C_{aa}C_{bb}
\label{ab_equil}
\end{equation}

As a result, the total energy minimized over $a\;$and $b$ can be written as 
\begin{equation}
E_{\min }=Ku^{2}  \label{En_equil}
\end{equation}

\begin{equation}
K=\frac{%
C_{bb}(C_{ua}^{2}-4C_{aa}C_{uu})+C_{ab}(C_{ab}C_{uu}-C_{ua}C_{ub})+C_{aa}C_{ub}^{2}%
}{\Delta }  \label{spring_equil}
\end{equation}

\bigskip We can now see that six effective spring constants $C_{\alpha \beta
}$ define the equilibrium slope parameters $a$ and $b~$and the equilibrium
energy $E_{\min }.$

The effective spring constants can be found from Eq. \ref{Spring_c} .
Sometimes, however, it is more practical to use the energy values $%
E[\{u,a,b\}]\;$defined for different sets $\{u,a,b\}$ of the boundary
parameters.\newline

For every distance $d$ \ between the insertions, the elastic energy $%
E[\{u,a,b\}]\;$can be calculated numerically \cite{JorMilPar2002}. Then, all
six spring constants can be found. The following equations illustrate this
approach:%
\begin{eqnarray*}
C_{uu} &=&\frac{E[\{u,0,0\}]}{u^{2}},\;C_{aa}=\frac{E[\{0,a,0\}]}{a^{2}}%
,\;C_{bb}=\frac{E[\{0,0,b\}]}{b^{2}}, \\
C_{ua} &=&\frac{E[\{u,a,0\}]-C_{uu}u^{2}-C_{aa}a^{2}}{ua};C_{ub}=\frac{%
E[\{u,0,b\}]-C_{uu}u^{2}-C_{bb}b^{2}}{ub}; \\
C_{ab} &=&\frac{E[\{0,a,b\}]-C_{aa}a^{2}-C_{bb}b^{2}}{ab};
\end{eqnarray*}%
\newline
With these constants, the equilibrium (or minimized over the slope
parameters) interaction energy profile $E_{\min }(d)$ can be determined from
Eq. \ref{Eq_2_same}$.\;$

\bigskip

\section{Final remarks}

It was shown that interaction energy between the insertions can be described
in terms of effective spring constants accounting for the coupling between
various degrees of freedom introduced through the boundary conditions. After
the spring constants are defined, the equilibrium slope (which can in
general become anisotropic) and corresponding interaction energy can be
defined analytically. This approach is much more efficient than the direct
energy minimization used in \cite{JorMilPar2002}. Some applications of this
approach will be considered elsewhere.

\bibliographystyle{aip}
\bibliography{D://Data/Word/LaTex/ref1}
%,D://Data/Word/LaTex/ref2,D://Data/Word/LaTex/ref3,D://Data/Word/LaTex/PRL_footnotes}

\bigskip

\end{document}